\documentclass[aps,prd,twocolumn,showpacs,nofootinbib]{revtex4-1}

\usepackage{amssymb} \usepackage{amsmath} \usepackage{graphicx}
\usepackage{epsfig,latexsym}

\RequirePackage{xspace} \allowdisplaybreaks

\begin{document}

\def\bef{\begin{figure}}
\def\eef{\end{figure}}

\newcommand{\nl}{\nonumber\\}

\newcommand{\ans}{ansatz }
\newcommand{\be}[1]{\begin{equation}\label{#1}}
\newcommand{\beq}{\begin{equation}}
\newcommand{\ee}{\end{equation}}
\newcommand{\beqn}[1]{\begin{eqnarray}\label{#1}}
\newcommand{\eeqn}{\end{eqnarray}}
\newcommand{\bd}{\begin{displaymath}}
\newcommand{\ed}{\end{displaymath}}
\newcommand{\mat}[4]{\left(\begin{array}{cc}{#1}&{#2}\\{#3}&{#4}
\end{array}\right)}
\newcommand{\matr}[9]{\left(\begin{array}{ccc}{#1}&{#2}&{#3}\\
{#4}&{#5}&{#6}\\{#7}&{#8}&{#9}\end{array}\right)}
\newcommand{\matrr}[6]{\left(\begin{array}{cc}{#1}&{#2}\\
{#3}&{#4}\\{#5}&{#6}\end{array}\right)}
\newcommand{\cvb}[3]{#1^{#2}_{#3}}
\def\lsim{\raise0.3ex\hbox{$\;<$\kern-0.75em\raise-1.1ex
e\hbox{$\sim\;$}}}
\def\gsim{\raise0.3ex\hbox{$\;>$\kern-0.75em\raise-1.1ex
\hbox{$\sim\;$}}}
\def\abs#1{\left| #1\right|}
\def\simlt{\mathrel{\lower2.5pt\vbox{\lineskip=0pt\baselineskip=0pt
           \hbox{$<$}\hbox{$\sim$}}}}
\def\simgt{\mathrel{\lower2.5pt\vbox{\lineskip=0pt\baselineskip=0pt
           \hbox{$>$}\hbox{$\sim$}}}}
\def\unity{{\hbox{1\kern-.8mm l}}}
\newcommand{\eps}{\varepsilon}
\def\ep{\epsilon}
\def\ga{\gamma}
\def\Ga{\Gamma}
\def\om{\omega}
\def\omp{{\omega^\prime}}
\def\Om{\Omega}
\def\la{\lambda}
\def\La{\Lambda}
\def\al{\alpha}
\newcommand{\ov}{\overline}
\renewcommand{\to}{\rightarrow}
\renewcommand{\vec}[1]{\mathbf{#1}}
\newcommand{\vect}[1]{\mbox{\boldmath$#1$}}
\def\tm{{\widetilde{m}}}
\def\mcirc{{\stackrel{o}{m}}}
\newcommand{\Dm}{\Delta m}
\newcommand{\dm}{\varepsilon}
\newcommand{\tanb}{\tan\beta}
\newcommand{\nbar}{\tilde{n}}
\newcommand\PM[1]{\begin{pmatrix}#1\end{pmatrix}}
\newcommand{\up}{\uparrow}
\newcommand{\down}{\downarrow}
\def\omE{\omega_{\rm Ter}}
%

\newcommand{\Dsusy}{{susy \hspace{-9.4pt} \slash}\;}
\newcommand{\DCP}{{CP \hspace{-7.4pt} \slash}\;}
\newcommand{\mc}{\mathcal}
\newcommand{\gr}{\mathbf}
\renewcommand{\to}{\rightarrow}
\newcommand{\gtc}{\mathfrak}
\newcommand{\wh}{\widehat}
\newcommand{\br}{\langle}
\newcommand{\kt}{\rangle}


\def\lsim{\mathrel{\mathop  {\hbox{\lower0.5ex\hbox{$\sim$}
\kern-0.8em\lower-0.7ex\hbox{$<$}}}}}
\def\gsim{\mathrel{\mathop  {\hbox{\lower0.5ex\hbox{$\sim$}
\kern-0.8em\lower-0.7ex\hbox{$>$}}}}}

\def\nn{\\  \nonumber}
\def\de{\partial}
\def\brf{{\mathbf f}}
\def\bbf{\bar{\bf f}}
\def\bF{{\bf F}}
\def\bbF{\bar{\bf F}}
\def\bA{{\mathbf A}}
\def\bB{{\mathbf B}}
\def\bG{{\mathbf G}}
\def\bI{{\mathbf I}}
\def\bM{{\mathbf M}}
\def\bY{{\mathbf Y}}
\def\bX{{\mathbf X}}
\def\bS{{\mathbf S}}
\def\bb{{\mathbf b}}
\def\bh{{\mathbf h}}
\def\bg{{\mathbf g}}
\def\bla{{\mathbf \la}}
\def\bmu{\mathbf m }
\def\by{{\mathbf y}}
\def\bmu{\mbox{\boldmath $\mu$} }
\def\bsig{\mbox{\boldmath $\sigma$} }
\def\bunity{{\mathbf 1}}
\def\cA{{\cal A}}
\def\cB{{\cal B}}
\def\cC{{\cal C}}
\def\cD{{\cal D}}
\def\cF{{\cal F}}
\def\cG{{\cal G}}
\def\cH{{\cal H}}
\def\cI{{\cal I}}
\def\cL{{\cal L}}
\def\cN{{\cal N}}
\def\cM{{\cal M}}
\def\cO{{\cal O}}
\def\cR{{\cal R}}
\def\cS{{\cal S}}
\def\cT{{\cal T}}
\def\eV{{\rm eV}}
%

\title{Super-light baryphotons, Weak Gravity Conjecture and Exotic instantons in Neutron-Antineutron transitions}

\author{Andrea Addazi$^1$}\email{3209728351@qq.com}
\affiliation{$^1$ Center for Field Theory and Particle Physics \& Department of Physics, Fudan University, 200433 Shanghai, China}

\begin{abstract}

In companion papers \cite{Addazi:2015pia,Addazi:2016rgo}, we have discussed current  
bounds  
of a new super-light baryo-photon, associated to a $U(1)_{B-L}$ gauged, 
from neutron-antineutron current data, which are competitive with E\"otv\"os type experiments. 
Here, we discuss the implications of a possible baryo-photon detection in string theory 
and quantum gravity. The discovery of a very light gauge boson 
should imply the violation of the
Weak Gravity Conjecture, carrying deep consequences 
in our understanding of holography, quantum gravity and black holes. 
On the other hand, we show how 
the detection of a baryo-photon would also exclude the generation 
of all $B{-}L$ violating operators from Exotic Stringy Instantons. 
We will disclaim the common statement in literature that neutron-antineutron may indirectly test 
at least a $300-1000\, \rm TeV$ scale. 
Searches of baryo-photons can provide indirect informations of the
Planck (or String) scale (quantum black holes, holography and non-perturbative stringy effects). 
This strongly motivates new neutron-antineutron experiments with adjustable magnetic fields
dedicated to the detection of super-light baryo-photons.

\end{abstract}

\maketitle

\section{Introduction}

As it is known,  $B$ and $L$ are
accidental symmetries
of the Standard Model. Their conservation is in agreement 
with all current data.  
Some symmetry principles could be behind $B,L$ accidental conservations. 
The simplest idea is to recover 
$B,L$ number conservations as
residual discrete symmetries of spontaneously broken 
 global $U(1)_{L}$, $U(1)_{B}$,  or a linear combination of the two 
 (as $U(1)_{B-L}$ or $U(1)_{B+L}$ and so on). 
 This class of models predicts the existence of new pseudo-goldstone
 bosons known in literature as Majorons
\cite{CM,SV} \footnote{Majorons can also provide a good candidate of (warm) dark matter 
\cite{Berezinsky:1993fm}. See also Refs.\cite{Berezhiani:1992cd,Berezhiani:1994jc,Kachelriess:2000qc,Kachelriess:2000qc,Lattanzi:2007ux,Lattanzi:2014zla,Boucenna:2014uma}.}.

An alternative way is to gauge $B,L$ symmetries. 
However, as it is well known, $U(1)_{B}$ and $U(1)_{L}$ gauged 
are anomalous. 
The only 
way-out from anomalies 
is to consider a $U(1)_{\zeta(B-L)}$ gauged,
where $\zeta$ is an arbitrary constant
which can be redefined in particle charges, 
i.e. $U(1)_{B-L}$ for convention. 
In particular, $U(1)_{B-L}$ gauge group may be spontaneously  broken by  
a new Higgs singlet field (Higgs mechanism) or a Stueckelberg gauged axion 
(Stueckelberg mechanism). 
Usually, $U(1)_{B-L}$ is thought as 
a spontaneously broken gauge group at 
high scales, i.e. a new $Z'$ boson possibly 
testable at LHC or future colliders. 
On the other hand, from the point of view 
of quantum field theory consistency, 
a gauge $U(1)_{B-L}$
could also be massless. 
But certainly, this would be not phenomenologically healthy: it would 
be in contradiction with baryogenesis
which necessary requests a violation of $B-L$
\footnote{See Ref.\cite{McKeen:2015cuz} for recent considerations on Baryogenesis 
and $n\bar{n}$ transitions. }. 
If we desired to break $B-L$ at a intermediate 
or high scale while a semi-massless gauge boson, we would introduce a very weakly coupled $B-L$
boson with $M_{b}^{2}\sim \alpha_{B-L}v_{B-L}^{2}$
and $\alpha_{B-L}<<1$. Such a scenario would be {\it technically natural}: a so tiny coupling remains stable against 
renormalization gauge group corrections.
In fact, all corrections in the Renormalization Group Equation (RGE)
are controlled by an overall factor $g_{B-L}^{3}$.
(as it can be understood counting two loop corrections in 
Landau gauge). 
For example the two-loop  RGE (in Landau gauge) contributions are suppressed 
as $g_{B-L}^{3}{\rm Tr}[Y^{\dagger}Y]$ (from Yukawa's couplings $Y$)
and  $g_{B-L}^{3}g_{i}^{2}$ (from gauge fields $g_{i}$). 
For instance, this is not the case of $U(1)_{B}$ or $U(1)_{L}$ gauged
which would be corrected by quadratically divergent contributions
and they should be enormously fine-tuned from their mass scale 
to the Planck scale.   
However, the new Higgs field $\chi$ introduced to spontaneously break $U(1)_{B-L}$
can mix with the ordinary Higgs field as $\chi^{\dagger}\chi \bar{H}H$ and this could introduce a new hierarchy 
problem. This is connected with the old hierarchy problem of the Higgs mass,
which presently remains still unsolved. 

\begin{figure}[htb] \label{BARI}
\begin{center}
\caption{Neutron-Antineutron transition in a baryo-photon background $\langle b_{0} \rangle$. The presence 
of a baryo-photon background field generates a mass splitting among neutron and antineutron. 
A Majorana mass term for the neutron can be generated by the a spontaneous symmetry breaking 
of $U(1)_{B-L}$ induced by the VEV of $\chi$.  
  }
\includegraphics[scale=0.085]{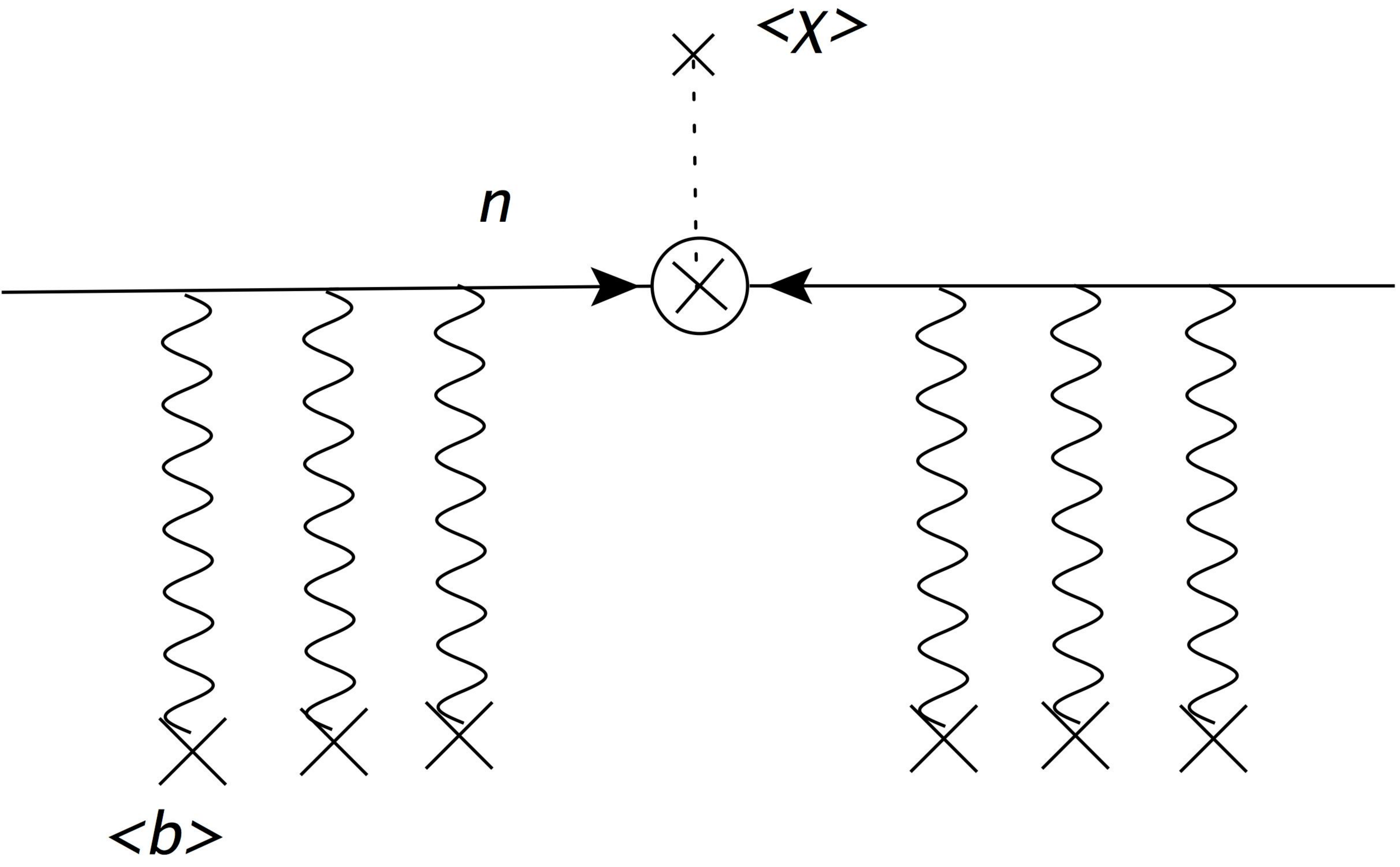}  
\end{center}
\vspace{-1mm}
\end{figure}

\begin{figure}[htb] \label{MDA2}
\begin{center}
\caption{a) The mixed disk amplitude coupling the physical RH (super)quark $U$ with two instantonic zero modes
$\tau$ and $\alpha$. In (b) the Mixed disk amplitude dual picture in terms of intersecting D-branes. 
 a) The mixed disk amplitude coupling the physical RH (super)quark $D$ with two instantonic zero modes
$\tau$ and $\beta$. In (d) the Mixed disk amplitude dual picture in terms of intersecting D-branes.  }
\includegraphics[scale=0.08]{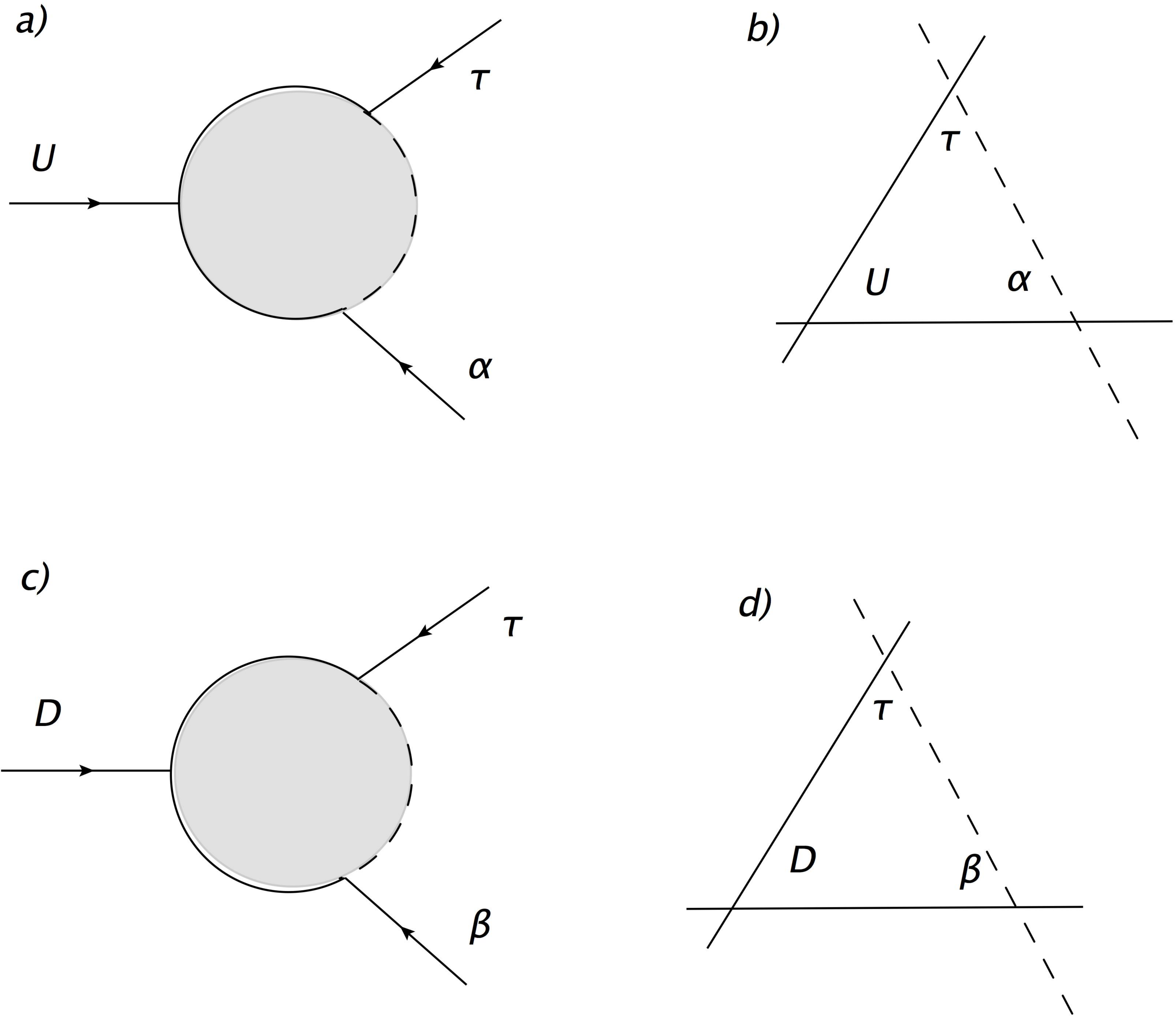}
 \end{center}
\vspace{-1mm}
\end{figure}

On the other hand, not all possible allowed gauge interactions in quantum field theories decoupled by gravity
seem to be compatible with quantum gravity. 
For instance, 
the Weak gravity conjecture (WGC) states that the weakest interaction is gravity
and it excludes the presence of new very light $U(1)$ bosons 
like $U(1)_{B-L}$ coupled to ordinary matter.
This means that for each interactions, it must exist 
a particle satisfying 
\be{mq}
\frac{m}{q}\leq M_{Pl}
\ee
where $m,q$ are mass and U(1)-charge of the particle respectively \cite{ArkaniHamed:2006dz}. 

At the present status, WGC is only based on heuristic arguments 
sustained by
holography and absence of global symmetries in quantum gravity and string theory.
In particular, L. Susskind suggested that, according to holography, Black Hole remnants should be impossible \cite{Susskind:1995da}. 
The WGC argument is the following: 
let us consider an hypothetical interaction of a particle with mass $m$
and $\tilde{\alpha}<1$, where $\tilde{\alpha}=\alpha_{YM}/G_{N}m^{2}$.
 In this case 
a black hole can have a charge from $0$ to $\bar{Q}=\tilde{\alpha}^{-1}$ (for example $\tilde{\alpha}\sim 10^{-10}$, i.e. $\bar{Q}\sim 10^{10}$)
and these charges cannot be radiated away 
as Hawking's radiation. This should imply a final remnant extremal 
BH with $M=QM_{Pl}$ and $Q$ in range $(0,\bar{Q}]$, contradicting Susskind's arguments. 
This seems to lead to the conclusion that WGC is sustained from the holographic principle.

One could think
that a heuristic argument 
may be not enough satisfiable
and 
the conjecture should be tested with high precision. 
To test WGC should be crucially important for our understanding of 
quantum gravity, holography and black holes.
For instance, a violation of WGC would imply that some fundamental aspect in our understanding of black holes
and quantum gravity is still missing. 
In particular, it is commonly retained that holography
is a crucial feature of black hole physics
and a violation of WGC could lead to revisit such a  concept itself. 

However, the detection of super-light baryo-photon 
can lead to rethink semiclassical non-perturbative solutions
in string theory. In particular, exotic D-brane instantons 
can generate $B{-}L$ violating operators
and their implications in particle physics 
were recently discussed in our papers 
\cite{Addazi:2014ila,Addazi:2015ata,Addazi:2015rwa,Addazi:2015hka,Addazi:2015eca,Addazi:2015fua,Addazi:2015oba,Addazi:2015goa,Addazi:2015yna,Addazi:2016xuh,Addazi:2016mtn}. 
 As is known, $B{-}L$ violating exotic instantons have necessary to be synchronized 
 with a Stueckelberg mechanism for $U(1)_{B-L}$, sending the 
 $B-L$ gauge boson mass to a large scale. 
 So that, a super-light baryo-photon is in tension with exotic instanton effects,
 which should be suppressed by non-perturbative stringy corrections
 beyond the semiclassical approximation. 
So that, the detection of a baryo-photon implies a prohibition 
of exotic instanton effects from the spontaneous symmetry breaking scale 
$v_{B-L}$ up to the String scale!

In this letter, we suggest to test both the weak gravity conjecture 
and non-perturbative stringy effects 
 from neutron-antineutron 
oscillations data. The neutron-antineutron transition was not observed
and last limits were obtained in by Baldo-Coelin et al. \cite{BC}.
From these data, Z. Berezhiani, Y. Kamyshkov and the author of this paper
have recently discussed limits 
to a new bary-photons coupled to the (anti)neutron from neutron-antineutron experiments 
\cite{Addazi:2015pia,Addazi:2016rgo}. 
The possibility to improve current neutron-antineutron limits was discussed in Ref.\cite{Phillips:2014fgb}.
However, authors of Ref.\cite{Phillips:2014fgb} \footnote{
See also related discussion on perturbative renormalization of $n\bar{n}$ operators \cite{Buchoff:2015qwa}
and on experimental aspects \cite{Davis:2016uyk}.} have emphasized aspects of 
 neutron-antineutron 
experiments as a test for the effective $\Delta B=2$ Majorana mass operator 
$(udd)^{2}/M^{5}$, in order to indirectly test $M\simeq 1000\, \rm TeV$ scale. 
So that, it was suggested to search $n-\bar{n}$ transitions with very suppressed external magnetic field 
$(B<10^{-4}\, \rm Gauss)$. But according to our papers \cite{Addazi:2015pia,Addazi:2016rgo}, 
a neutron-antineutron transition should be suppressed by the presence of 
an external baryo-photon background field. For example, for a baryo-photon 
background field with scale $10^{-11}\div 10^{-13}\, \rm eV$ on the Earth surface, 
neutron-antineutron transitions would be enhanced in strong magnetic field conditions
$B\sim 1\div 10\, {\rm Gauss}$ rather than suppressed ones. 
In this letter, we suggest 
that the search of a baryo-photon can provide 
a test for the Planck (and String) scale physics, 
even if $M_{Pl},M_{s}>>1000\, \rm TeV$.
In fact, according to 
our considerations above, 
the detection of a baryo-photon in neutron-antineutron
should violate Weak Gravity Conjecture 
as well as should be a test for exotic D-brane instantons.
In other words, a detection of a baryo-photon would lead to re-discuss 
the same basic principles of quantum gravity and string theory, 
such as holography, stringy instantons, black hole remnants and so on.
In this sense, searches for bary-photons in neutron-antineutron experiments can indirectly test quantum gravity.

\section{Baryo-photon }

The Baryo-photon model is based on
Standard Model gauge group extension with 
an extra $B-L$ gauge symmetry: 
$SU(3)_{c}\times SU(2)_{L}\times U(1)_{Y} \times U(1)_{B-L}$. 

The $B-L$ baryo-photon gauge coupling with neutron, proton and lepton currents is 
\be{Int}
\mathcal{L}_{B-L}=g b_{\mu}(\bar{n}\gamma^{\mu}n+\bar{p}\gamma^{\mu}p-\bar{e}\gamma^{\mu}e-\bar{\nu}\gamma^{\mu}\nu)
\ee
where $b_{\mu}$ is the baryo-photon associated with $U(1)_{B-L}$.
With an exact $U(1)_{B-L}$, the neutron-antineutron transition is forbidden, 
otherwise a gauge symmetry is unlikely violated. 
So that, $U(1)_{B-L}$ has to be spontaneously broken
and this can be synchronized with the generation of a effective Majorana mass for the neutron.
For example, one can introduce effective operators
like
\be{effective}
\frac{\chi}{M^{6}}(udd)(udd),\,\,\,\,\frac{\chi}{M^{6}}(qqd)(qqd),\,\,\,\,\frac{\chi}{M^{6}}(udd)(qqd)
\ee
$(qq=\epsilon^{\alpha\beta}q_{\alpha}q_{\beta}/2=u_{L}d_{L})$
where $\chi$ is a Higgs scalar field with charge $Q_{B-L}=-2$
and getting a vacuum expectation value 
$\langle \chi \rangle =v_{\chi}$. 
In this case, a $n-\bar{n}$ transition is generated 
with an effective suppression scale $\mathcal{M}_{n\bar{n}}=(M^{6}/v_{\chi})^{1/5}$
and consequently a Majorana mass term $\delta m_{n\bar{n}}=\Lambda_{QCD}^{6}/\mathcal{M}_{n\bar{n}}$.
An example of UV completion of such an operator 
was suggested in Refs. \cite{Berezhiani:2015afa,Addazi:2016rgo} as a see-saw mechanism for the neutron. 
Alternatively, it is also possible that the generation of the effective Majorana 
mass term for the neutron is totally disconnected by the 
spontaneous symmetry breaking mechanism 
and it happens after the spontaneous breaking.
Then, in full generality, one can also consider the more complicated case
in which $U(1)_{B-L}$ is spontaneously broken
by a combination of scalars $\chi,\eta_{i}$:
\be{Mb}
v=[v_{\chi}^{2}+(Q_{i}/Q_{\chi})^{2}]^{1/2},\,\,\,M_{b}=2^{3/2}gv
\ee
where $Q_{i},Q_{\chi}$ are B-L charges of the scalars
while $M_{b}$ is the mass of the baryophoton. 
As a consequence, the baryophoton mediates 
a spin-indpendent force among SM particles with baryon and lepton charges:
\be{VBL}
V_{i}=\alpha_{B-L}\frac{Q_{i}Q_{A}}{r}e^{-r/\lambda},\,\,\,\lambda =M_{b}^{-1}
\ee
$\alpha_{B-L}=g^{2}/4\pi$ and where 
$$\lambda\simeq 0.6\times (10^{-49}/\alpha_{B-L})^{1/2}(1\,\rm keV/v)\times 10^{16}\, {\rm cm}$$

So that, an external $B-L$ static background generates an effective 
mass splitting term among the neutron $(Q_{B-L}=+1)$ and the antineutron $(Q_{B-L}=-1)$:
\be{Vnnar}
\frac{V_{n\bar{n}}}{V_{n}^{G}}=\pm \tilde{\alpha}q_{A}
\ee
where $V_{n}^{G}$ is the gravitational potential,
$q_{A}=Q_{A}m_{n}/(M_{A})$
and $\tilde{\alpha}=\alpha_{B-L}/\alpha_{G}$
and $\alpha_{G}=G_{N}m_{n}^{2}$.
If a $\tilde{\alpha}<<1$ gauge boson was detected, 
WGC would be violated. 

Yukawa radius larger than Earth's radius
\be{lambda}
\lambda>R_{Earth}\rightarrow \alpha_{B-L}<10^{-49},\,\,\,\tilde{\alpha}<1.7\times 10^{-11}
\ee

The Earth induces a gravitational energy for the neutron at its radius 
$V_{Earth}^{E}=-Gm_{n}M_{Earth}/R_{Earth}\simeq 0.66\,\rm eV$,
while the Sun $V_{Sun}^{G}=-Gm_{n}M_{Sun}/AU\simeq 10\, \rm eV$,
while the Galaxy $V_{Galaxy}^{G}\simeq 1\, \rm keV$.
The total energy potential contribution from baryo-photon on a (anti)neutron in laboratory frame is 
\be{Vn}
V_{n}=\tilde{\alpha}(0.5V_{Earth}^{G}e^{-R_{Earth}/\lambda}+0.13V_{Sun}^{G}e^{-AU/\lambda}
\ee
$$+0.13V_{Galaxy}^{G}e^{-10\, \rm kpc/\lambda})$$

The effective interaction enters in the oscillation probability as 
\be{Pnnbar}
P_{n\bar{n}}=P^{+}+P^{-}
\ee
$$P^{\pm}=\frac{\delta m_{n\bar{n}}^{2}}{{\delta m_{n\bar{n}}^{2}+\Delta_{\pm}^{2}}}\sin^{2}\left(t \sqrt{\delta m_{n\bar{n}}^{2}+\Delta_{\pm}^{2}} \right)$$
where $\Delta_{\pm}=V\mp \Omega_{B}$, 
$\Omega_{B}=|{\bf \mu}_{n}\cdot {\bf B}|\simeq 6 \cdot 10^{-12}{\rm (B/1G)eV}$
(Zeeman energy shift induced by the external magnetic field),
$\pm$ corresponds to two polarizations states,
$\delta m_{n\bar{n}}$ is the effective Majorana mass term.

In Fig. 3, we report various exclusion plots for $(\lambda,\alpha_{B-L})$ parameter space
compared with E\"otv\"os-like experiments. 
As one can see, for $\lambda>10^{9}\, \rm cm$, which is comparable with the Earth radius, 
for $v_{B-L}>1\, \rm GeV$ the parameter space is very constrained. 
On the other hand, $v_{B-L}< 1\, \rm meV$ is not possible in a minimal 
model: it would imply a spontaneous breaking of $U(1)_{B-L}$ only in very late Universe
$(1\div 10\, \rm Gyrs)$ which is clearly excluded by baryogenesis. 
However, we suggest that this scenario could have a subtle way-out:
it is possible that $B-L$ was broken in the early Universe because of thermal bath 
induced expectation values to one (or more) scalar Higgs,
allowing Baryogenesis, and restored later.
For instance, our idea is inspired by various old models of high temperature symmetry breaking 
 suggested in Refs.\cite{R1,R1b,R2,R3,R4,R5}. 
An interesting possibility could be a phase transition  
 mechanism from a electroweak conserving and $B-L$ broken vacua $(G_{SM})$
 to an electroweak breaking and $B-L$ preserving vacua $G'=(SU(3)_{c}\times U(1)_{em}\times U(1)_{B-L})$.
 In this case, CP-violating scatterings of primordial plasma to expanding Bubbles 
 associated to the broken-restored phase $G'$ can
 generate a Baryon-asymmetry as in standard electroweak baryogenesis (See \cite{Morrissey:2012db} for a review).

Among the landscape of parameters, we would like 
to point out the attention on $v_{B-L}\simeq 1\, \rm keV$
allowing for $\lambda \simeq 10^{16}\, \rm cm$
a $\Delta V=|V_{n}-V_{\bar{n}}|\simeq 10^{-11}\, \rm eV$
which would correspond to a magnetic field of $5\, \rm Gauss$
($10$ times the Earth magnetic field or so)
coupled to the neutron magnetic moment. 
As a consequence, a so strong background would completely 
suppress a $n-\bar{n}$ transition searched in condition 
of $|B|<10^{-4}\, \rm Gauss$ as suggested in Ref. \cite{Phillips:2014fgb}.
On the contrary in this case a neutron-antineutron transition should be searched
in resonant condition $|\mu_{n}\cdot B|\simeq \Delta V$. 
Roughly speaking, neutron-antineutron experiments
seriously risk to not detect any new physics with the wrong magnetic field set-up.

\begin{figure}[htb] \label{MDA}
\begin{center}
\caption{The parameter space of $({\rm log}_{10}\,\lambda(cm),\,{\rm log}_{10}\, \alpha_{B-L})$ 
is constrained by E\"otv\"os type experiments, as displayed in this figure (in green, Adelberg (2012))
(we applied limits discussed in Ref.\cite{A12} for a B-L baryo-photon). 
We display the range from $\Delta \lambda=10^{9}\div 10^{23}\, {\rm cm}$
and $\Delta \alpha_{B-L}=10^{-42} \div 10^{-56}$. 
We report several different excluded regions for various values of VEV $v_{B-L}$ (from $1\, \rm meV$ to $1\, \rm GeV$)
(the region down the black lines is excluded by neutron-antineutron data).
(With $G$ we label the Galaxy range scale). See also Fig.1-2 of Ref. \cite{Addazi:2016rgo}. }
\includegraphics[scale=0.08]{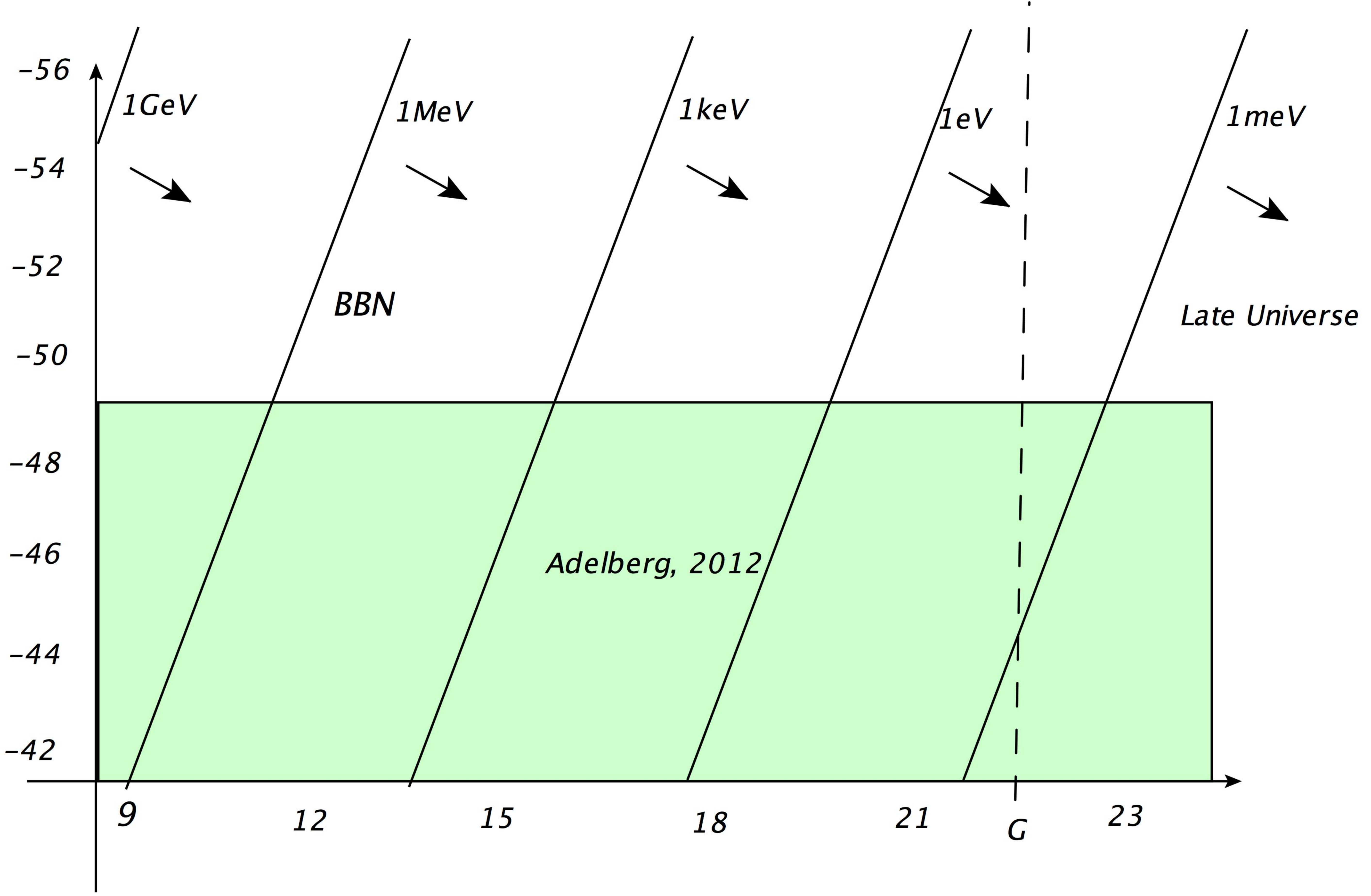}  
\end{center}
\vspace{-1mm}
\end{figure}

\section{Exotic instantons}

 The possible detection of a so light bary-photons would also rule-out B-L violating exotic instantons.
 In Refs. \cite{Addazi:2015goa}, we have shown how the intersection of E2-branes, wrapping different 3-cycles on $CY_{3}$,
 with D6-brane stacks can generate new non-perturbative neutron-antineutron operators.
 For instance, the effective lagrangian is 
 \be{effla}
 \mathcal{L}_{E2}=c^{(1)}_{ff'f''}\tau_{i,f}U_{f'}^{i}\alpha_{f''}+c^{(2)}_{f}\tau_{i,f'''}D_{f^{IV}}^{i}\beta_{f^{V}}
 \ee
 where $\tau,\alpha,\beta$ are chiral fermionic zero modes (or modulini) associated to the 
 Exotic instanton, while $U,D$ are RH up and down quarks. 
 In Fig.2, we report the mixed disk amplitudes generating the effective lagrangian 
 Eq.(\ref{effla}) from string theory. 
 Integrating over the modulini space, 
 \be{UDD}
 \mathcal{W}=\frac{\mathcal{Y}_{f_{1}f_{2}f_{3}f_{4}f_{5}f_{6}}e^{-S_{E2}}}{M_{S}^{3}}U_{f_{1}}D_{f_{2}}D_{f_{3}}U_{f_{4}}D_{f_{5}}D_{f_{6}}
 \ee
 where $\mathcal{Y}$ is a $3\times 6$ flavor matrix, combination of $c^{(1),(2)}$ couplings.  
The same lagrangian Eq.(\ref{effla}) can be considered 
with a one-half reduce number of modulini families,
providing a trilinear $\Delta B=1$ term 
 \be{UDD}
 \mathcal{W}=y_{f_{1}f_{2}f_{3}}e^{-S_{E2'}}U_{f_{1}}D_{f_{2}}D_{f_{3}}
 \ee
 This operator can generate a Neutron-Antineutron transition 
 mediated by a gluino exchange connecting quark-squark reduction currents. 
There are several different exotic instanton solution which 
cannot preserve $U(1)_{B-L}$ even if not directly connected 
to $n-\bar{n}$ transitions. 
For example exotic instantons with an effective lagrangian 
\be{effn}
\mathcal{L}_{E2'}=k^{(1)}_{ff'f''}N_{f}\alpha_{f'} \beta_{f''} 
\ee
 that integrating on the modulini space
 generates a Majorana mass matrix for the RH neutrino
 \be{MajoranM}
 \mathcal{W}_{E2}=M_{S}e^{-S_{E2''}}NN
 \ee
 As is well known, such an operator can generate 
 a Majorana mass for the LH neutrino 
 from a see-saw type I mechanism. 
 Alternatively, a Weinberg superpotential 
$\mathcal{W}=e^{-S_{E2^{'''}}}HLHL/M_{S}$
 can be directly generated
 by 
 \be{HLHL}
 \mathcal{L}=h_{1}\gamma_{\alpha}  L^{\alpha} \delta+h_{2}\gamma'_{\alpha}H^{\alpha}\delta'
 \ee

 However, the generation of these superpotential is incompatible with a B-L light baryphotons. 
 In fact, the generation of $n-\bar{n}$ is necessary synchronized
 with a Stueckelberg mechanism of $U(1)_{B-L}$. 
 In fact, all the $e^{-S_{E2}}$ factors have a structure
 \be{SE2}
 e^{-S_{E2}}=e^{-\mathcal{V}_{\Pi}/g_{s}+i\sum_{r}c_{r}a_{r}}
 \ee
 where $\mathcal{V}_{\Pi}$ is the volume of $\Pi$-cycles wrapped by a E-brane
 on the internal $CY$;
$a_{r}$ are RR axions and $c_{r}$ are E-brane couplings to them, 
$g_{s}$ is the string-coupling constant associated to the vacuum expectation value of the dilaton field
($g_{s}=e^{\langle \phi \rangle}$). 
The Eq.(\ref{SE2}) is not invariant under RR axion shifts, i.e. under $U(1)_{B-L}$ in our case: 

\be{SE2}
e^{-S_{E2}}\rightarrow e^{-i\sum_{A}N_{A}(I_{MA}-I_{MA^{*}})\Lambda_{A} }e^{-S_{E2}}
\ee
where $I$ is the umber of intersection among the E-brane $M$ and the background D-brane $M$,
$N_{A}$ is the number of A D-brane stacks and $\Lambda$ is an axion shift constant. 
 \footnote{See Refs. 
\cite{Ibanez:2006da,Blumenhagen:2006xt,Ibanez:2007rs,Blumenhagen:2009qh,Blumenhagen:2007bn,Blumenhagen:2007zk,Cvetic:2010mm,Abe:2015uma,Cvetic:2009yh}
  for more details on these aspects.}. 
and as is known this is exactly compensated by the shift factor of the superpotential combinations.
The shift is associated to a Stueckelberg mechanism for B-L. 
 As a consequence, the associated B-L boson gets a huge mass,
 typically of the order of the string scale or so.

 So that, we can argue that the observation of a very light baryo-photon
 would have strong implications for string phenomenology. 
 In fact, this could imply that a {\it non-perturbative protection mechanism}
would suppress all possible B-L exotic instantons for many orders 
 magnitude from the string scale to the low scale of B-L spontaneous symmetry breaking.  
 For instance, effects of RR and NS-NS fluxes wrapped by Euclidean D-branes
 could strongly suppress the mixed disk amplitudes associated to exotic instantons.

\section{Conclusions and remarks}  
  
In this letter, we discussed possible implications of the detection 
of a super-light bary-photon coupled to (anti)neutrons 
in quantum gravity and string theory.
Current available measures of $n-\bar{n}$ experiments 
impose unexpectedly stringent 
bounds to the 
baryo-photon mass and coupling constant. 
We have discussed how the detection of a 
super-weak baryo-photon may rule out the Weak Gravity Conjecture
as well as the generation of non-perturbative $(B{-}L)$-violating operators  
from exotic D-brane instantons. 
It is commonly retained that neutron-antineutron 
experiments would indirectly test at least 
$1000\, \rm TeV$ scale physics in next generation of experiments \cite{Phillips:2014fgb}. 
However, we want to disclaim such a statement. 
In fact, following our arguments, neutron-antineutron experiments could
 indirectly test the Weak Gravity Conjecture 
with very high precision.

We also have stressed how the detection of the super-weak baryo-photon 
may rule out the presence of $B-L$ violating Exotic Stringy Instantons 
up to the String scale!
 In fact, Exotic Instantons must 
necessary be associated with a Stueckelberg mechanism, 
providing a large mass to the $B-L$ gauge field. 
In other words, 
a so light baryo-photon 
should be {\it sequestered} by
Exotic Instantons, generating, for example, a mass 
term for the neutrino, or other R-parity violating operators.

\vspace{1cm}

{\large \bf Acknowledgments} 
\vspace{1mm}

I acknowledge enlightening 
discussions with  Massimo Bianchi and Zurab Berezhiani.

\end{document}